\documentclass{article}

\usepackage{cite}
\usepackage{graphicx}
\usepackage{amsmath,amssymb}

\def\GR{general relativity}
\def\cy{cylindrical}
\def\cyl{cylindrically symmetric}

\def\wh{wormhole}
\def\whs{wormholes}
\def\asflat{asymptotically flat}

\def\Jl#1#2{{\it #1} {\bf #2},\ }
\def\CQG#1 {\Jl{Class. Quantum Grav.}{#1}}
\def\GC#1 {\Jl{Grav. Cosmol.}{#1}}
\def\GRG#1 {\Jl{Gen. Rel. Grav.}{#1}}
\def\PRD#1 {\Jl{Phys. Rev. D}{#1}}
\def\PRL#1 {\Jl{Phys. Rev. Lett.}{#1}}

\def\cm{\hspace*{1cm}}

\def\eq{Eq.\,}

\def\d{\partial}
\def\beq{\begin{equation}}
\def\eeq{\end{equation}}
\def\bear{\begin{eqnarray}}
\def\bearr{\begin{eqnarray} &&}
\def\ear{\end{eqnarray}}
\def\earn{\nonumber \end{eqnarray}}

\def\half{{\tfrac{1}{2}}}
\def\sign{\mathop{\rm sign}\nolimits}
\def\diag{\mathop{\rm diag}\nolimits}
\def\const{{\rm const}}

\def\R{{\mathbb R}}
\def\kappa{\varkappa}
\def\e{{\rm e}}

\def\mn{_{\mu\nu}}
\def\MN{^{\mu\nu}}
\def\mN{_\mu^\nu}

\begin{document}
\begin{center}
{\large\bf Rotating cylindrical wormholes: a no-go theorem}\\[5mm]

K.A. Bronnikov\footnote
	{E-mail: kb20@yandex.ru}
\\[5mm]
{\small\it Center for Gravitation and Fundam. Metrology, VNIIMS,
	   	Ozyornaya St. 46, Moscow 119361, Russia;\\
	Institute of Gravitation and Cosmology, PFUR,
		Miklukho-Maklaya St. 6, Moscow 117198, Russia;\\
	National Research Nuclear University ``MEPhI''
		(Moscow Engineering Physics Institute),\\
		Kashirskoe sh. 31, Moscow 115409, Russia
		}
\bigskip

\parbox{14cm}
  {The existing solutions to the Einstein equations describing rotating
  cylindrical wormholes are not asymptotically flat and therefore cannot
  describe wormhole entrances as local objects in our Universe. To overcome
  this difficulty, flat asymptotic regions are added to wormhole solutions
  by matching them at some surfaces $\Sigma_-$ and $\Sigma_+$. It is shown,
  however, that if the wormhole solution is obtained for scalar fields with
  arbitrary potentials, possibly interacting with an azimuthal electric or
  magnetic field, then the matter content of one or both thin shells
  appearing on $\Sigma_-$ and $\Sigma_+$ violate the Null Energy Condition.
  Thus exotic matter is still necessary for obtaining a twice asymptotically
  flat wormhole.  }

\end{center}
\bigskip

  Traversable Lorentzian wormholes are widely discussed in gravitational
  physics since they lead to many effects of interest like time machines
  or shortcuts between distant parts of space. Large enough \whs, if any,
  can lead to observable effects in astronomy \cite{sha, accr, kir-sa1}.

  In attempts to build realistic \wh\ models, the main difficulty is that
  in \GR\ (GR) and some of its extensions a static \wh\ geometry requires
  the presence of ``exotic'', or phantom matter, that is, matter violating
  the weak and null energy condition (WEC and NEC), at least near the
  throat, the narrowest place in a \wh\ \cite{thorne,viss-book,HV97,ws_book}.
  This result is obtained if the throat is a compact 2D surface with a
  finite area \cite{HV97}.

  Examples of phantom-free \wh\ solutions have been obtained in extensions
  of GR, such as the Einstein-Cartan theory \cite{BGal15},
  Einstein-Gauss-Bonnet gravity \cite{GBo}, brane worlds \cite{BKim03}, etc.
  We here prefer to consider GR as a theory quite well describing the
  reality on the macroscopic scale while the extensions more likely concern
  very large densities and/or curvatures. In GR there are phantom-free \wh\
  models with axial symmetry, including the Zipoy \cite{zipoy} and
  superextremal Kerr vacuum solutions as well as solutions with scalar and
  electromagnetic fields possibly interacting with each other
  \cite{br-fab97, matos15}; in all of them, however, a disk that plays the
  role of a \wh\ throat is bounded by a ring singularity whose existence may
  be though of as an unpleasant price paid for the absence of exotic matter.

  The above-mentioned result of \cite{HV97} does not directly apply to
  objects like cosmic strings, infinitely stretched along a certain
  direction, in the simplest case \cyl\ ones. Such systems with and
  without rotation were discussed, in particular, in
  \cite{BLem09,BLem13,BS14,BK15} (see also references therein). It was shown
  there, with a number of examples, that phantom-free \cy\ \wh\ solutions to
  the Einstein equations are rather easily obtained.

  A problem with \cy\ systems is their undesirable asymptotic behavior.
  To describe \whs\ potentially visible to distant observers like
  ourselves in our very weakly curved Universe, one has to require their
  flat (or string\footnote
	{Or at least \asflat\ up to an angular deficit well-known in
	cosmic string configurations. In what follows, for brevity, we
	speak of asymptotic flatness though cosmic-string asymptotics
	could also be mentioned on equal grounds.})
  asymptotics, which is hard to achieve under \cy\ symmetry.

  To overcome this difficulty, it was suggested \cite{BLem13} to build \whs\
  with flat asymptotic regions on both sides of the throat by matching a
  \wh\ solution to suitable parts of Minkowski space-time, but no successful
  examples of phantom-free \whs\ were so far obtained in this way.
  In the present paper we further discuss this opportunity and prove a no-go
  theorem on the conditions under which this cut-and-paste procedure cannot
  lead to \asflat\ phantom-free \whs.

  Consider a stationary \cyl\ metric of the form
\beq                                                    \label{ds-rot}
       ds^2 = \e^{2\gamma(u)}[ dt - E(u)\e^{-2\gamma(u)}\, d\varphi ]^2
       - \e^{2\alpha(u)}du^2 - \e^{2\mu(u)}dz^2 - \e^{2\beta(u)}d\varphi^2,
\eeq
  where $u$, $z$ and $\varphi$ are the radial, longitudinal and angular
  coordinates. This metric is said to describe a \wh\ if either (i) the
  circular radius $r(u) = \e^{\beta(u)}$ has a minimum (called an
  $r$-throat) and is large (preferably tends to infinity) far from this
  minimum or (ii) a similar behavior is observed for the area function $a(u)
  = \e^{\mu+\beta}$ (its minimum is called an $a$-throat) \cite{BLem09,
  BLem13}. If a \wh\ is \asflat\ at both extremes of the $u$ range, it
  evidently possesses both kinds of throats.

  The vortex gravitational field existing in a space-time with the metric
  (\ref{ds-rot}) can be characterized by the angular velocity $\omega(u)$
  given by \cite{BLem13, kr2, kr4}
\beq                                                  \label{om}
     \omega = \half (E\e^{-2\gamma})' \e^{\gamma-\beta-\alpha}.
\eeq
  under an arbitrary choice of the coordinate $u$ (a prime stands for
  $d/du$). Furthermore, the reference frame comoving to a matter distribution
  in its motion by the angle $\varphi$ is determined by a zero component
  $T^3_0$ of the stress-energy tensor (SET), hence (via the Einstein
  equations) by the Ricci tensor component $R_0^3 \sim (\omega
  \e^{2\gamma+\mu})' = 0$, and thus in this reference frame we have
\beq       	      					\label{omega}
	\omega = \omega_0 \e^{-\mu-2\gamma}, \cm \omega_0 = \const.
\eeq

  It then turns out \cite{BLem13} that the diagonal components of the Ricci
  ($R\mN$) and Einstein ($G\mN = R\mN - \half \delta\mN R$) tensors split
  into the corresponding components for the static metric (that is,
  (\ref{ds-rot}) with $E=0$) plus the $\omega$-dependent addition, so that
  $G\mN = {}_s G\mN + {}_\omega G\mN$, where ${}_s G\mN$ is the static part,
  and
\beq
	{}_\omega G\mN = \omega^2 \diag (-3,\ 1,\ -1,\ 1).  \label{Ein-o}
\eeq
  Moreover, the tensors ${}_s G\mN$ and ${}_\omega G\mN$ (each separately)
  satisfy the conservation law $\nabla_\alpha G^\alpha_\mu =0$ in this
  static metric. Hence, by the Einstein equations $G\mN = - \kappa T\mN$,
  the tensor ${}_\omega G\mN/\kappa$ acts as an additional SET with exotic
  properties (e.g., the effective energy density is $-3\omega^2/\kappa <
  0$), making it easier to satisfy the existence conditions for both $r$-
  and $a$-throats, as confirmed by a number of examples in
  \cite{kr4,BLem13,BK15}.

  In all these examples, however, the \whs\ are not \asflat; even more
  than that, as is clear from \eq (\ref{omega}), any solution obtained in
  this way cannot be \asflat\ since it would require $\omega \to 0$ along
  with finite limits of $\gamma$ and $\mu$, which is incompatible with
  (\ref{omega}).

  The following trick was suggested \cite{BLem13} for obtaining an \asflat\
  configuration: to cut a non-\asflat\ \wh\ configuration at some regular
  cylinders $\Sigma_+\,(u=u_+)$ and $\Sigma_-\,(u=u_-)$ on different sides
  of the throat and to match it there with flat-space regions extending to
  infinity. Then the junction surfaces comprise thin shells with certain
  surface SETs, and it remains to check whether or not these SETs satisfy 
  the WEC and NEC.

  We will show, however, that with a large class of matter sources of the
  \wh\ solution, it is impossible to obtain the surface SETs on both
  $\Sigma_+$ and $\Sigma_-$ respecting the NEC.

  Indeed, if we choose the harmonic radial coordinate $u$ \cite{kb79a,kb79b}
  specified by the relation $\alpha = \beta + \gamma + \mu$, then a certain
  combination of the Einstein equations takes the form
\beq                                                            \label{03}
	\beta'' - \gamma'' - 4\omega_0^2 \e^{2\beta-2\gamma}
		= \kappa \e^{2\alpha} (T^t_t - T^\varphi_\varphi).
\eeq
  Therefore, if the SET of matter satisfies the condition $T^t_t =
  T^\varphi_\varphi$, \eq (\ref{03}) is easily integrated giving
\beq
	\e^{\beta-\gamma} = \frac{1}{2|\omega_0| s(k,u)}, 	\label{bg}
	     \ \ \ \ k = \const,
\eeq
  where one more integration constants has been excluded by choosing
  the origin of the $u$ coordinate, and the function $s(k, u)$ is defined as
\beq                                                          \label{def-s}
	s(k,u) =  \left\{
		\begin{array}{ll}
	k^{-1} \sinh ku, & \ \  k > 0,\ \ u \in \R_+;\\
		      u, & \ \ k=0, \ \ u \in \R_+; \\
	 k^{-1} \sin ku, & \ \ k<0, \ \ 0 < u < \pi/|k|.
		\end{array}\right.
\eeq

  The condition $T^t_t = T^\varphi_\varphi$ holds for a large class
  of matter Lagrangians in the metric (\ref{ds-rot}), such as, e.g.,
\beq                                                        \label{L_se}
	L = g\MN \d_\mu\phi\d_\nu\phi - 2V(\phi) - P(\phi)F\MN F\mn,
\eeq
  with an arbitrary scalar field potential $V(\phi)$ and an arbitrary
  function $P(\phi)$ characterizing the scalar-electromagnetic interaction,
  assuming that $\phi = \phi(u)$ and that the Maxwell tensor $F\mn$
  describes a stationary azimuthal magnetic field ($F_{21}=-F_{12}\ne 0$) or
  its electric analogue.

  Quite a general observation follows from \eq (\ref{bg}):
  in the case $k < 0$, if the remaining field equations give the metric 
  functions $\mu$ and $\alpha$ which are finite and regular on the
  segment $0 \leq u \leq \pi/|k|$ (or, equivalently, on any other half-wave
  of the function $\sin ku$), then the whole configuration is a \wh, with
  both $r$- and $a$-throats; both $r(u)$ and $a(u)$ tend to infinity as
  $u \to 0$ and $u \to \pi/|k|$, but these limiting surfaces are singular 
  due to $\e^\gamma \to 0$ and $\omega \to \infty$. Configurations with 
  $k \geq 0$ can also be of \wh\ nature, as is confirmed by examples of 
  vacuum and scalar-vacuum solutions \cite{BLem13,BK15}.

  Now suppose we have such a \wh\ metric and cut it at some regular points
  $u=u_-$ to the left and $u= u_+$ to the right of both $r$- and
  $a$-throats. Let us try to join it at $u=u_\pm$ to regions of Minkowski
  space-time with the metric $ds_{\rm M}^2 = dt^2 - dx^2 - dz^2 - x^2
  d\varphi^2$ outside suitable cylinders $x = x_\pm = \const$. Assuming a
  rotating reference frame with an angular velocity $\Omega = \const$, we
  substitute $\varphi \to \varphi + \Omega t$ to obtain
\beq                                                          \label{ds_M}
      ds_{\rm M}^2 = dx^2 + dz^2 + x^2 (d\varphi + \Omega dt)^2 - dt^2.
\eeq
  The relevant quantities in the notations of (\ref{ds-rot}) are
\beq                                                   \label{M-param}
      \e^{2\gamma} =  1 - \Omega^2 x^2,\qquad
      \e^{2\beta} = \frac{x^2}{1 - \Omega^2 x^2},\qquad
      E = \Omega x^2, \qquad
      \omega = \frac{\Omega}{1 - \Omega^2 x^2}.
\eeq
  This metric is stationary and can be matched to an internal metric at
  $|x| <  1/|\Omega|$, inside the ``light cylinder'' on which the linear
  rotational velocity reaches that of light.

  Matching of two \cyl\ regions at a surface $\Sigma:\ u=u_0$ means that
  we identify the two metrics on this surface, so that
\beq                                                       \label{ju-1}
      [\beta] = 0, \qquad [\mu] = 0, \qquad
      [\gamma] = 0, \qquad [E] =0,
\eeq
  where, as usual, the brackets denote discontinuities: for any $f(u)$, $[f]
  = f(u_0+0) - f(u_0 -0)$. One should note that in general the metrics on
  different sides of the surface $\Sigma$ may be written using different
  choices of the radial coordinate $u$, but it does not matter since the
  quantities involved in all matching conditions used are insensitive to the
  choice of $u$.

  The next step is to determine the material content of the junction
  surface $\Sigma$ according to the Darmois-Israel formalism
  \cite{israel-67,BKT-87}: in our case of a timelike surface $\Sigma$, the
  surface SET $S_a^b$ is expressed in terms of $K_a^b$, the extrinsic
  curvature of $\Sigma$, as
\def\tK {{\tilde K}{}}
\beq                                                        \label{ju-2}
	S_a^b = - \kappa^{-1} [\tK_a^b], \qquad
		\tK_a^b := K_a^b - \delta_a^b K^c_c,
		\qquad a, b, c = 0, 2, 3.
\eeq

  Assume that the matching conditions (\ref{ju-1}) on $\Sigma_\pm$,
  identifying the surfaces $x=x_\pm$ in Minkowski regions and $u=u_\pm$ in
  the internal region, are fulfilled by fixing the values of $x_\pm$ for
  given $u_\pm$ and other parameters of the system, including the values of
  $\Omega = \Omega_\pm$ in each Minkowski region. It is important that we
  should take $x_+ >0$ and $x_- < 0$ to adjust the directions of the normal
  vectors to $\Sigma_\pm$.

  Now, the question is whether the surface SETs on $\Sigma_\pm$ can satisfy
  the WEC whose requirements are
\beq                                                             \label{WEC}
	S_{00}/g_{00} = \sigma \geq 0, \qquad\
		S_{ab}\xi^a \xi^b \geq 0,
\eeq
  where $\xi^a$ is any null vector ($\xi^a \xi_a =0$) on $\Sigma=\Sigma_\pm$;
  the second inequality in (\ref{WEC}) comprises the NEC as part of the WEC.
  The conditions (\ref{WEC}) are equivalent to
\beq                                                            \label{WEC1}
      [\tK_{44}/g_{44}] \leq 0, \qquad\ [K_{ab}\xi^a \xi^b] \leq 0.
\eeq
  If we choose two null vectors on $\Sigma$ in the $z$ and $\varphi$
  directions as
\beq                                                        \label{xi_1,2}
	\xi^a_{(1)} = (\e^{-\gamma},\ \e^{-\mu},\ 0),
\qquad\
    \xi^a_{(2)} = (\e^{-\gamma}+ E\e^{-\beta-2\gamma},\ 0,\ \e^{-\beta}),
\eeq
  the conditions (\ref{WEC1}) read \cite{BLem13}
\beq                                                       \label{WEC2}
       [\e^{-\alpha}(\beta'+\mu')] \leq 0,
    \qquad
       [\e^{-\alpha}(\mu' - \gamma')] \leq 0,
    \qquad
       [\e^{-\alpha}(\beta'-\gamma') + 2\omega] \leq 0.
\eeq
  Now we can apply these requirements to our configuration at both
  junctions. Consider the third condition which contains the function
  $\beta-\gamma$ given by Eq.\,(\ref{bg}).
  Using it, on $\Sigma_-$ with $x = x_- < 0$ we obtain
\beq                                                          \label{N2-}
       \e^{-\alpha (u_-)} \, \frac{(-s'+ \sign\omega_0)}{s}
       	      + \frac{(1+\Omega_- x)^2}{|x|(1-\Omega_-^2 x^2)} \leq 0,
\eeq
  and on $\Sigma_+$ with $x = x_+ > 0$ we have in a similar way
\beq                                                          \label{N2+}
       \e^{-\alpha(u_+)} \, \frac{(s'- \sign\omega_0)}{s}
       		+ \frac{(1+\Omega_+ x)^2}{x(1-\Omega_+^2 x^2)} \leq 0.
\eeq
  Here $s$ and $s'=ds/du$ refer to the function $s=s(k,u)$ introduced in
  (\ref{def-s}).

  The inequalities (\ref{N2-}) and (\ref{N2+}) lead to the conclusion that
  the NEC (hence also the WEC) cannot be satisfied on both $\Sigma_+$ and
  $\Sigma_-$ simultaneously.

  Indeed, assuming $\omega_0 > 0$, the inequality (\ref{N2-}) can only hold
  if $1 - s'(k,u) < 0$ at $u=u_-$. But $s'(k,u) = \{\cosh ku, \ 1,\
  \cos|k|u\}$ for $k >0,\ k=0$ and $k < 0$, respectively, and only at $k >
  0$ we have $1 - s' < 0$. Thus at $\Sigma_-$ the solution in the internal
  region should be taken with $k >0$. On the contrary, (\ref{N2+}) can hold
  only if $1 - s'(k,u) >0$ at $u=u_+$, which is only possible if $k <0$. But
  any particular solution describing the internal region has a fixed value
  of $k$, hence the inequalities (\ref{N2-}) and (\ref{N2+}) are
  incompatible with each other. Furthermore, if $\omega_0 < 0$, we have in
  the first term in (\ref{N2-}) $-s'-1 < 0$, making no problem; however, in
  (\ref{N2+}) there is, instead, $s'+1 > 0$, hence this inequality cannot
  hold whatever be the parameter $k$.

  We conclude that the NEC is inevitably violated at least on one of the
  surfaces $\Sigma_+$ and $\Sigma_-$.

  This general result has been obtained for any source of gravity in
  \wh\ solutions such that $T^t_t = T^\varphi_\varphi$ in the metric
  (\ref{ds-rot}), in particular, those given by (\ref{L_se}). Possible 
  solutions with other sources are certainly not ruled out. So the problem of
  obtaining potentially observable, stationary, nonsingular, phantom-free
  \whs\ in GR remains open.

\end{document}